\documentclass[fleqn,twoside]{article}
\usepackage{espcrc2,macros}
\font\sixrm=cmr6

\usepackage{epsfig}
\usepackage[figuresright]{rotating}

\newcommand{\be}{\begin{equation}}
\newcommand{\ee}{\end{equation}}
\newcommand{\bea}{\begin{eqnarray}}
\newcommand{\eea}{\end{eqnarray}}

\newcommand{\AmS}{{\protect\the\textfont2
  A\kern-.1667em\lower.5ex\hbox{M}\kern-.125emS}}

\title{
\begin{flushright}
{\small 
MS-TP-01-12\\
Bicocca-FT-01-21 \\
$$ \quad  $$ \\
}
\end{flushright} 
Quenched twisted mass QCD at small quark masses and in large 
volume\thanks{Based on a poster presented by R.~Frezzotti at the
XIX International Symposium on Lattice Field Theory "Lattice 2001", 
August 19-24, 2001, Berlin, Germany.}} 
\author{ M. Della Morte\address[BICO]{Universit\`a di Milano-Bicocca,  
                                  Dipartimento di Fisica,
         Piazza della Scienza 3, I-20126 Milano, Italy},
         R. Frezzotti\addressmark[BICO] 
         and
         J. Heitger\address{WWU M\"unster,
                 Institut f\"ur Theoretische Physik, 
         Wilhelm-Klemm-Str. 9, D-48149 M\"unster, Germany}
        }

\begin{document}

\begin{abstract}
As a test of quenched lattice twisted mass QCD, we compute the 
non-perturbatively O($a$) improved pseudoscalar and vector 
meson masses and the pseudoscalar decay constant down to 
$M_{\rm PS}/M_{\rm V} = 0.467(13)$ at $\beta=6$ in large volume.
We check the absence of exceptional configurations and -- by 
further data at $\beta=6.2$ -- the size of scaling violations. The
CPU time cost for reaching a given accuracy is close to that with ordinary
Wilson quarks at $M_{\rm PS}/M_{\rm V} \simeq 0.6$ and grows  
smoothly as $M_{\rm PS}/M_{\rm V}$ decreases.
\end{abstract}


\vspace{-2cm}
\maketitle

\section{INTRODUCTION}

The lattice twisted mass QCD (tmQCD) action \cite{paper1}
in infinite space-time volume reads
\bea \label{Wils_tm_act1}
  S_{\rm W}[U,\psibar,\psi] = S_{\rm g} + a^4\sum_{x}
  \psibar_x [D_{\rm tm} \; \psi]_x \; , 
 \nonumber \\
D_{\rm tm} = \gamma \wt{\nabla}
- \frac{a}{2} \nabla^\ast \nabla +m_0 +i\muq\gamma_5\tau^3 \; ,
\eea
where $\wt{\nabla} = (\nabla+\nabla^\star)/2$,
$\psi$ is a flavour quark doublet and the matrix $\tau^3$
acts in the flavour space. If $m_0 \equiv \mc$,
the quark mass is entirely given by the
{\em multiplicatively} renormalized parameter $\muq$. 

Lattice tmQCD is a regularization of QCD with $N_{\rm f}=2$
mass degenerate quarks \cite{paper1} where an action term that is proportional to $\muq$ 
and is not aligned in flavour chiral space with the Wilson term
contributes to the soft breaking of the chiral symmetry. 
As a consequence, the lowest eigenvalue of the 
Hermitean square of the Dirac matrix on any gauge background is bounded
from below -- in lattice units -- by $(a\muq)^2$. The problem of the exceptional
configurations in quenched and partially quenched QCD is thus solved 
as long as the quark mass is not vanishingly small.

One can think of the two mass-degenerate quark flavours as 
the $u$ and $d$ quarks in a chirally twisted basis,
with the obvious consequences on the physical meaning of 
the correlation functions \cite{paper1}. Heavier flavours of 
Wilson quarks can still be added e.g. in the usual way,
while the neglected tiny mass difference between $u$ and $d$ quarks can
safely be taken into account by means of chiral perturbation theory.  
Moreover, lattice tmQCD can also be formulated for a doublet of 
non-degenerate quarks, whilst retaining the protection against
exceptionals on all gauge backgrounds\cite{RFnotes}.  

\section{OBSERVABLES AND SIMULATIONS}
In this study we restrict attention to the pseudoscalar and vector meson
masses, $M_{\rm PS}$ and $M_{\rm V}$, the pseudoscalar decay constant,
$F_{\rm PS}$, and the renormalized quark masses, $\mur$ and $\mr$, which
are defined according to eq.~(3.3) and eq.~(3.20) of Ref.~\cite{paper3}.  
It is convenient for our discussion to define polar quark mass "coordinates" via
\be \label{polar_mass}
\tan \alpha = \mur/\mr \, , \quad
M_\rmR=(\mur^2 + \mr^2)^{1/2} \, .
\ee

%%%%%%%%%%%%%%%%%%%%%%%%%%%%%%%%%%%%%%%%%%%%%%%%%%%%%%%%%%%%%%%%%%%%%%%
\begin{table*} \label{tab1}
%\begin{table}[htb]
\begin{center}
\begin{tabular}{cccccccccc}
\hline
Set &  $\beta$  & $\kappa$, $\muq$ & $L/a$, $T/L$ & $\#$ meas. & $M_{\rm R} r_0$ & $M_{\rm PS}r_0$ & $F_{\rm PS}r_0$ & $M_{\rm V}r_0$ \\
\hline
A1    &    6      & 0.135196,   0.0266 &  16, 2 &  650 &  0.2729(15)  & 1.711(7) & 0.455(5) & 2.662(40)    \\
A1'   &    6      & 0.135196,   0.0266 &  16, 3 &  650 &  0.2729(15)  & 1.714(6) & 0.455(6) & 2.656(42)    \\
A2    &    6.2    & 0.135814,   0.0180 &  24, 2 &  535 &  0.2558(16)  & 1.623(8) & 0.456(5) & 2.557(32)    \\
B1L   &    6      & 0.135208,   0.0190 &  24, 2 &  260 &  0.1949(11)  & 1.452(6) & 0.432(6) & 2.517(35)    \\
B1    &    6      & 0.135208,   0.0190 &  16, 2 &  535 &  0.1949(11)  & 1.455(8) & 0.428(5) & 2.513(47)    \\
B2    &    6.2    & 0.135814,   0.0138 &  24, 2 &  300 &  0.1962(12)  & 1.420(9) & 0.436(7) & 2.462(41)    \\
C     &    6      & 0.135208,   0.0117 &  24, 2 &  260 &  0.1205(7)   & 1.160(6) & 0.401(6) & 2.485(59)    \\
\hline
\end{tabular}
\end{center}
\caption{}
\vspace{-20pt}
%\caption{Simulation parameters, statistics ($\#$ meas.) and results 
%for our observables.} 
%\end{table}
\end{table*}
%%%%%%%%%%%%%%%%%%%%%%%%%%%%%%%%%%%%%%%%%%%%%%%%%%%%%%%%%%%%%%%%%%%%%%%

We work in the {\em quenched} approximation considering systems of physical
size $L^3 T$ at $\beta=6$ and $\beta=6.2$, with $L$ such that $M_{\rm PS}L
\geq 4.5$ to suppress finite volume effects.
Following closely Ref.~\cite{paper3},
we adopt Schr\"odinger functional \makebox{(SF)} boundary conditions
for tmQCD and compute the SF correlators
\bea 
\fa^{11}(x_0)\, , \quad \fp^{11}(x_0)\, , \quad
\fv^{12}(x_0)\, , \quad f_1^{11} \, , 
\nonumber \\
\kv^{11}(x_0)\, , \quad \kt^{11}(x_0)\, ,\quad
\ka^{12}(x_0) \phantom{\, , \quad f_1^{11} \, ,}
\eea
%\fa^{11}(x_0)\, , \quad \fp^{11}(x_0)\, , \quad
%\fv^{12}(x_0)\, , \quad f_1^{11} \, , \nonumber \\
%\kv^{11}(x_0)\, , \quad \kt^{11}(x_0)\, ,\quad
%\ka^{12}(x_0) \phantom{\, , \quad f_1^{11} \, ,}
as a function of the Euclidean time $x_0$, for $L$ ranging from 1.5 to 
2.2~fm, depending on the quark mass values, and $T/L =$~2~or~3.
We hence construct the renormalized and O($a$) improved SF correlators
$[\fap^{11}]_{\hbox{\sixrm R}}(x_0)$, $[\fpp^{11}]_{\hbox{\sixrm R}}(x_0)$,
$[\kvp^{11}]_{\hbox{\sixrm R}}(x_0)$ and $[\ktp^{11}]_{\hbox{\sixrm R}}(x_0)$, 
which are specified in eq. (3.11) of Ref.~\cite{paper3} with $\alpha$ 
given by eq.~(\ref{polar_mass}).

In the limit of large $x_0$ and large $T-x_0$ and up to cutoff effects, 
the correlators
$[\fap^{11}(x_0)]_{\hbox{\sixrm R}}$ and $[\fpp^{11}(x_0)]_{\hbox{\sixrm R}}$,
which correspond to the insertion of the operators $(A'_{\rmR})^1_0$ and
$(P'_{\rmR})^1$, see eq.~(3.10) of Ref.~\cite{paper3}, are expected to be dominated
by the lowest isotriplet pseudoscalar state. In the same limit the correlators
$[\kvp^{11}(x_0)]_{\hbox{\sixrm R}}$ and $[\ktp^{11}(x_0)]_{\hbox{\sixrm R}}$,
which correspond to the insertion of the operators $(V'_{\rmR})^1_k$ and
$(T'_{\rmR})^1_{k0}$, see eq.~(3.10) of Ref.~\cite{paper3}, are dominated by
the lowest isotriplet vector state.

\begin{figure}[htb]
\vspace{-10pt}
\epsfig{file=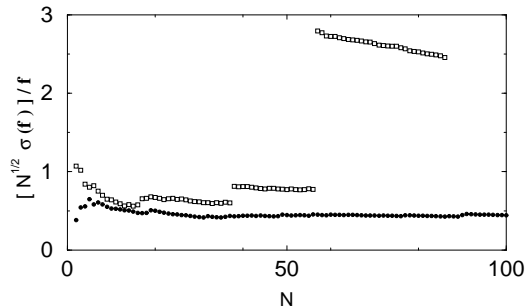,width=4.2cm,angle=-90} \\
\vspace{-33pt}
%%\framebox[55mm]{\rule[-21mm]{0mm}{43mm}}
\caption{The square root of the relative a priori variance of
$f=\fp^{11}(24a)$ versus the number of measurements $N$
for the simulation C (filled circles)
and a simulation with the same values of $\beta$ and 
$M_\rmR$, but $\muq=0$ (open squares).} 
\vspace{-15pt}
\label{fig_fP24}
\end{figure}

In our simulations we employ the non-perturbatively improved version
of the action~(\ref{Wils_tm_act1}), with $\csw$ taken from Ref.~\cite{O(a)2}.
An overview of our simulation parameters, statistics and preliminary results 
is given in Table~1, while $\alpha = \pi/2 + {\rm O}(a)$. 
%Due to O($a^2$) and \mbox{statistical} uncertainties
%on $\mc$, the angle $\alpha$ is not exactly $\pi/2$. 

\begin{figure}[htb]
\vspace{-10pt}
\epsfig{file=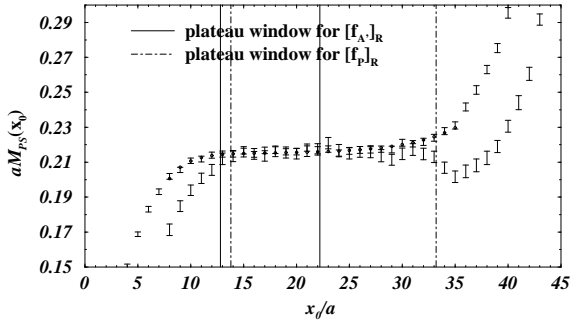,width=4.3cm,angle=-90} \\
\vspace{-33pt}
%%\framebox[55mm]{\rule[-21mm]{0mm}{43mm}}
\caption{Pseudoscalar effective masses extracted from the correlators
$[\fap^{11}]_\rmR$ and $[\fpp^{11}]_\rmR$ for data set C. The circles
denote our fit to $[\fap^{11}]_\rmR$, which accounts for O($a\muq$)
contributions from states with twisted parity and isospin at large $x_0$.}
\vspace{-15pt}
\label{fig_meff}
\end{figure}

Our most critical simulation (set C) required
$\sim 230$~GFlops~$\times$~day by employing
a CGNE solver for the SSOR-preconditioned Dirac matrix $D_{\rm tm}$. For
this (and similar) choice(s) of parameters the BiCG solver often does not
converge, while a CGNE solver works fine with small \mbox{fluctuations} in the
number of iterations. 

\section{RESULTS}

We find that lattice tmQCD
allows, as expected, to safely work in a region
of parameters which would be inaccessible
with ordinary Wilson quarks: see e.g. Fig.~\ref{fig_fP24}.
For a given number of independent measurements, the statistical errors on 
$M_{\rm PS}$ and $M_{\rm V}$ are comparable, up to a factor of one to three, 
to those  
found e.g. with domain wall quarks \cite{CPPACS_B_K}. The CPU time
effort, e.g. for the data sets A1, A1' and A2, is in line with 
the computational cost for ordinary Wilson quarks.

\begin{figure}[htb]
\vspace{-10pt}
\begin{center}
\epsfig{file=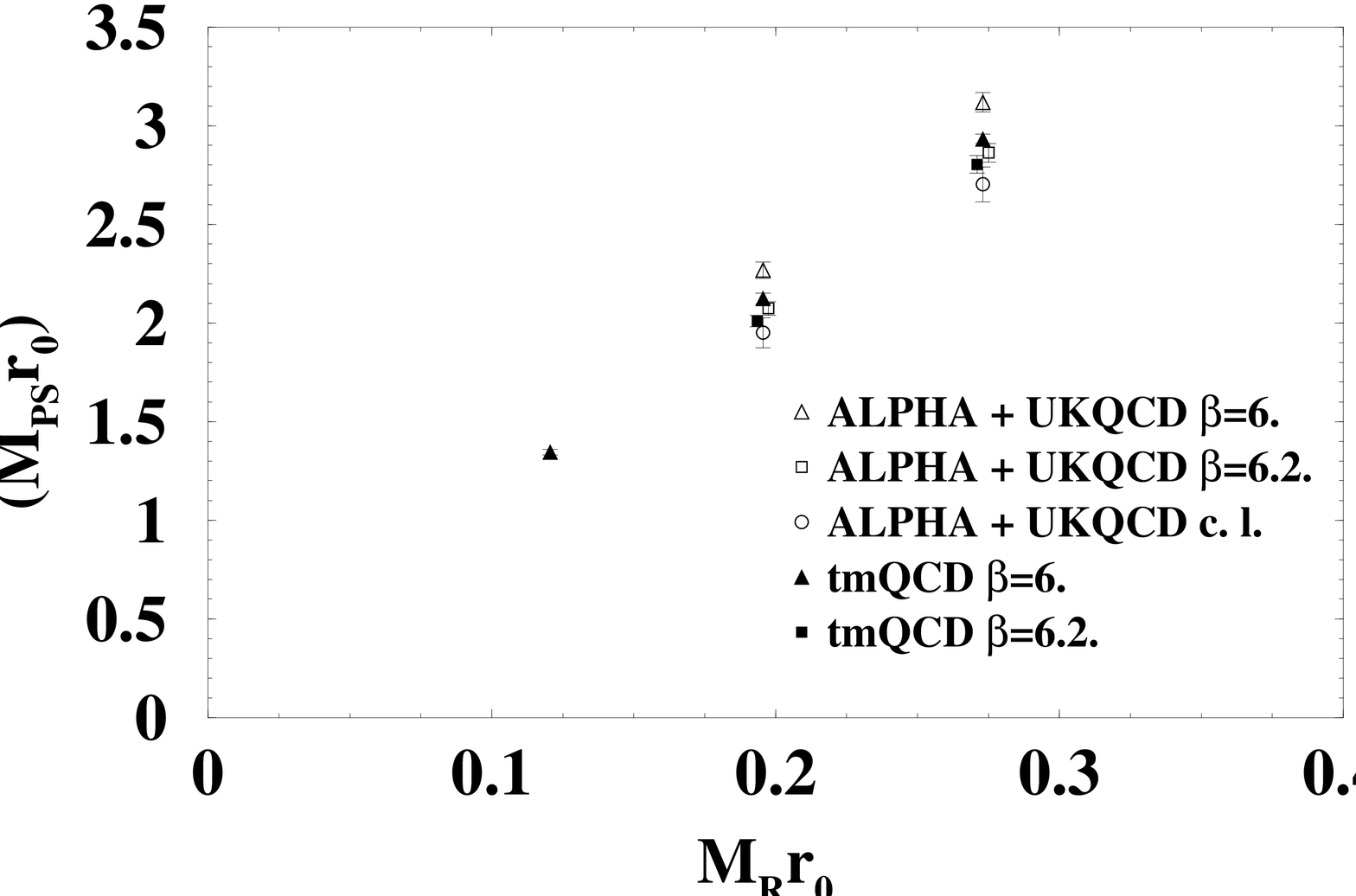,width=4.4cm,angle=-90} \\
\vspace{10pt}
\epsfig{file=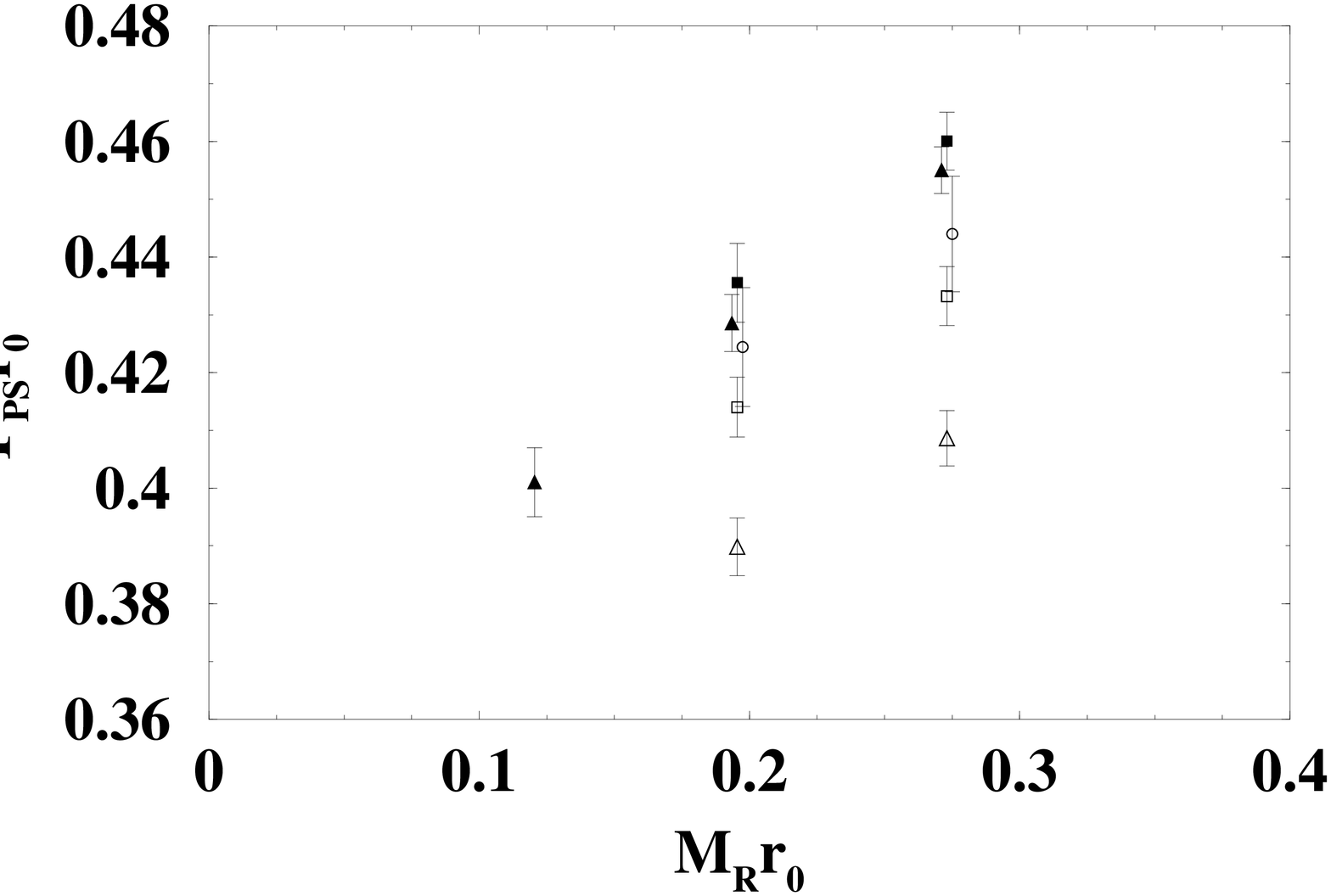,width=4.5cm,angle=-90} \\
%\vspace{15pt}
%%%\epsfig{file=Mrho_chi2.eps,width=3.724cm,angle=-90}
%\epsfig{file=MV_pro.eps,width=4.4cm,angle=-90}
\vspace{-28pt}
\end{center}
%%\framebox[55mm]{\rule[-21mm]{0mm}{43mm}}
\caption{$M_{\rm PS}r_0$ and $F_{\rm PS}r_0$ versus $M_{\rm R}r_0$ from
tmQCD and reanalysis of the data of Ref.~\cite{garden}, including a
continuum extrapolation (c. l.).} 
\vspace{-20pt}
\label{fig_chir}
\end{figure}

Concerning the extraction of hadron masses and matrix elements, we recall that
the lattice action~(\ref{Wils_tm_act1}) enjoys reduced parity and 
isospin symmetries \cite{paper1}: it just preserves
the third isospin generator and parity combined with a flavour 
exchange. At finite $a$
the correlators may hence receive contributions of order $a\muq$
from the allowed states with different parity and isospin. 
While deferring details to Ref.~\cite{paper4}, we
show in Fig.~\ref{fig_meff} an example of effective masses extracted
from SF correlators: the correlator $[\fap^{11}]_\rmR(x_0)$ 
receives contributions of order $a\muq$ that are peculiar to our SF
setup and would be absent at $\alpha=0$. 
Analogous effects in $[\fpp^{11}]_\rmR(x_0)$ are 
expected to be very small \cite{paper4}.
%No such effects are instead present,
%up to O($a^3$) terms, in $[\fpp^{11}]_\rmR(x_0)$.
 
As detailed in Ref.~\cite{paper3}, we expect the relations among
our observables and the renormalized parameters 
$r_0$ and $M_\rmR$ to be O($a$) improved.
In particular, when working at $\alpha = \pi/2 + {\rm O}(a)$,     
an O($a$) improved estimate of $F_{\rm PS}$
is obtained in terms of $M_{\rm PS}$
and the matrix element of e.g. the operator $2M_\rmR (P'_\rmR)^1$ between the
charged pion and the vacuum\cite{FSW_latt01}. 
%charged pion and the vacuum\footnote{
%For more details, see: R.~Frezzotti, S.~Sint and P.~Weisz, these proceedings. 
%}. 
This approach to $F_{\rm PS}$, which
is also valid in unquenched lattice tmQCD, requires to know no other
counterterms than $\zg$, $\mc$ and $\csw$. 

In order to check for the residual scaling violations, we produced data
at $\beta=6.2$ (sets A2 and B2), while keeping $\alpha$, $M_\rmR$ 
and $r_0$ fixed. More precisely, the ratio $\mr/M_\rmR$ was 
tuned\footnote{The small mismatch in $M_\rmR r_0$ for the set A2
was corrected by employing estimates of the dependence
of our observables on $M_\rmR r_0$.}
to the value $0.000(1)$ for $M_\rmR r_0 = 0.195(1)$ and
$-0.015(1)$ for $M_\rmR r_0 = 0.273(2)$. For the simulation C
we found $\mr/M_\rmR=0.083(5)$.
We also reanalysed the data of Ref.~\cite{garden}, which were
produced at $\beta=6,6.1,6.2,6.45$, by imposing precisely the same 
renormalization conditions as in this study of tmQCD. We then
performed a continuum extrapolation, assuming a purely quadratic
dependence on $(a/r_0)^2$ and discarding the data at $\beta=6$.
However, owing to the use of the one-loop estimate of $\ba$, the
resulting estimate of $F_{\rm PS}$ is not fully O($a$) improved.
The outcome of this exercise is compared with the results from tmQCD
in Fig.~\ref{fig_chir}. We omit the case of $M_{\rm V} r_0$, as
cutoff effects are hardly visible within statistical errors.  

%Based on these results, we expect that our lattice tmQCD approach can  
%straightforwardly and conveniently be applied to unquenched QCD.   

\section{CONCLUSIONS}

Lattice tmQCD solves the problem of exceptional configurations
and allows for rather precise results with a moderate computational
effort and tiny scaling violations, at least for the observables of
this study. The framework is suitable for extension
beyond the quenched approximation. 

This work is part of the ALPHA Collaboration research programme.

\end{document}